# Phase Behavior of Polyelectrolyte Block Copolymers in Mixed Solvents

Galder Cristobal[1], Jean-François Berret[2,@], Cedrick Chevallier[3], Ruela Talingting-Pabalan[3], Mathieu Joanicot[1] and Isabelle Grillo[4]

[1] : LOF, unité mixte CNRS/Rhodia/Bordeaux-I, 178 avenue du Dr Schweitzer, 33608 Pessac, France
[2] : Matière et Systèmes Complexes, UMR 7057 CNRS/Université Denis Diderot, Bâtiment Condorcet, 10 rue Alice Domon et Léonie Duquet, 75205 Paris, France
[3] : Complex Fluid Laboratory, UMR CNRS/Rhodia 166, CRTB Rhodia Inc., 350 George Patterson Blvd., Bristol, PA 19007 USA.
[4] : Institut Laue-Langevin, Large Scale Structures, 6 rue Jules Horowitz, B.P. 156, 38042 Grenoble, France

**Abstract :** We have studied the phase behavior of the poly(n-butyl acrylate)-b-poly(acrylic acid) block copolymer in a mixture of two miscible solvents, water and tetrahydrofuran (THF). The techniques used to examine the different polymers, structures and phases formed in mixed solvents were static and dynamic light scattering, small-angle neutron scattering, nuclear magnetic resonance and fluorescence microscopy. By lowering the water/THF mixing ratio X, the sequence unimers – micron-sized droplets – polymeric micelles was observed. The transition between unimers and the micron-sized droplets occurred at X ~ 0.75, whereas the microstructuration into core-shell polymeric micelles was effective below X ~ 0.4. At intermediate mixing ratios, a coexistence between the micron-sized droplets and the polymeric micelles was observed. Combining the different aforementioned techniques, it was concluded that the droplet dispersion resulted from a solvent partitioning that was induced by the hydrophobic blocks. Comparison of poly(n-butyl acrylate) homopolymers and poly(n-butyl acrylate)-b-poly(acrylic acid) block copolymers suggested that the droplets were rich in THF and concentrated in copolymers and that they were stabilized by the hydrophilic poly(acrylic acid) moieties.

## I – Introduction

During the last decade, block copolymers in solutions and in melts have been the subject of intensive experimental and theoretical investigations [1]. In solvents which are selective for one of the two blocks, copolymers self-assemble into micellar-type structures. Depending on the molar mass ratio between the two blocks and on the formulation, these structures can be locally spherical, cylindrical or lamellar. For spherical morphologies, the formation of the micelles have been explained theoretically as arising from the balance between the interfacial tension of the hydrophobic blocks in contact with water and the stretching of the blocks in the corona [2,3]. Microstructural studies on charged systems *i.e.* where the soluble part is a polyelectrolyte chain were also undertaken. Because of the electrostatic interactions between charged monomers, these blocks tend to stretch significantly, modifying then the energy balance previously mentioned and in fine the microstructure of the core-shell colloid [4].

Among the copolymers with a polyelectrolyte hydrosoluble blocks, poly(styrene)-b-poly(acrylic acid) (PS-b-PAA) was certainly the most studied system [5-18]. In water (a selective solvent for poly(styrene)), it was reported that PS-b-PAA form star-like aggregates comprising a dense core made of the insoluble PS and surrounded by an outer PAA shell [7,8]. The extension of the PAA chains was found to be highly sensitive to the degree of ionization of the acrylic acid monomers and to the ionic strength of the solution. At pH higher than the pKa of PAA (pKa = 5.5), the ionization of the chains is complete and the outer chains are almost fully stretched, giving rise to a morphology that resembles that of an urchin [19,20]. In PS-b-PAA, because the glass transition temperature of polystyrene is above room temperature ($T_G$ ca. 50° C [16,17,21]), the cores of the micelles are frozen and as such, it is supposed that there is no unimer exchange between the self-assembled structures.

In order to favor this unimer exchange, and eventually to be able to control aggregation, various strategies have been proposed. The first strategy consisted to utilize an hydrophobic block with a glass transition temperature lower than room temperature. Among different systems envisioned so far [1], poly(n-butyl acrylate)-b-poly(acrylic acid), hereafter referred as to PBA-b-PAA has appeared as an interesting candidate, since its $T_G$ is around - 52 ° C [13,22-28]. In systems where the dynamics of the blocks remain liquid-like, it was anticipated that the structures formed would be at thermodynamic equilibrium. Recent reports have shown contradictory results for PBA-b-PAA, in particular about the critical micellar concentration (cmc) in water [22,26,27]. Gaillard et al. have evaluated cmc's for PBA-b-PAA with molecular weights comprised between 2000 and 6000 g mol$^{-1}$ to be in the range $10^{-1}$ – $10^{-3}$ wt. % [22], whereas Théodoly and coworkers found for comparable molecular weights values lower than $10^{-4}$ wt. % [26]. Extremely low cmc's were suggested in these latter report, based on the measurements of the interfacial tension for PBA in water of the order of 20 mN m$^{-1}$.

Another strategy to modulate the aggregation morphologies was to dissolve the copolymers in an organic solvent that is common for both blocks, such as dimethylformamide (DMF), tetrahydrofuran (THF) or dioxane. These organic solvents have another remarkable property : they are miscible with water in all proportions [29]. Controlled self-assembly could be then induced starting from the unimer state, by a slow addition of water. This approach was suggested and investigated in great details by Eisenberg and coworkers during the last decade [5,6,9,30]. It was later extended to different copolymer systems [13,14,18,23,31-33]. Typical experiments conducted by Eisenberg et al. consisted in measuring the turbidity of a solution prepared in an organic solvent as a function of the water content. In PS310-b-PAA52 for instance (the indices here denote the number of repeating units of each block), Eisenberg found that with increasing water content, the turbidity of the solution exhibited a series of transitions which were indicative of a change in morphology [6,30]. These new aggregation states were observed by TEM which displayed various sequences, including spheres, rods, vesicles, lamellae and other intermediate structures [5,6,9,10,30,32,33]. According to these authors, the first turbidity transition showing up with the addition of water was related to the critical water concentration, noted cwc. Typical cwc values ranged between 1 - 20 %, depending on the organic solvent and on the molecular weight of the copolymer.

In the present paper, we have used poly(n-butyl acrylate)-b-poly(acrylic acid) copolymers because of the low value of the h-PBA glass transition temperature, and we have followed the aforementioned mixed solvent approach. The organic solvent was tetrahydrofuran (THF) and the phase behavior of symmetrical and asymmetrical PBA-b-PAA copolymers were investigated as a function of the mixing ratio between these two solvents. In this study, it was first verified that in water the copolymers self-assemble spontaneously into core-shell micelles. We have also demonstrated that the same copolymers were fully soluble in THF. The techniques used to analyze the different polymers, structures and phases formed were light and neutron scattering, NMR and fluorescence microscopy. As a result, we have found that the PBA-b-PAA copolymers in mixed THF/water solvents do display a cwc with addition of water (at about 25 % of added water), but the cwc was associated with a microphase separation and not with a morphology transition. The microphase separated state was described as a dispersion of micron-sized droplets, for which a partitioning of the two solvents was evidenced. To the best of our knowledge, such solvent partitioning has not been not reported in block copolymer systems. The phase behavior for h-PBA homopolymers were obtained for comparison, and confirmed the features observed with the diblocks.

## II – Experimental
### II.1 – Polymer Synthesis, Sample Preparation.

Poly(n-butyl acrylate)-b-poly(acrylic acid) block copolymers were synthesized by Madix® radical polymerization [34,35]. Developed at Rhodia, the Madix® technology used the xanthate as the chain-transfer agent in the controlled radical polymerization. The synthesis was performed in ethanol in two steps, the hydrophobic blocks being first polymerized and the hydrophilic blocks being added to the previous blocks in a second step. Poly(n-butyl acrylate) homopolymers (h-PBA) were extracted from the first polymerization step aliquots, whereas poly(acrylic acid) homopolymers (h-PAA) were synthesized separately. The chemical formulae of the monomers are shown in Fig. 1.

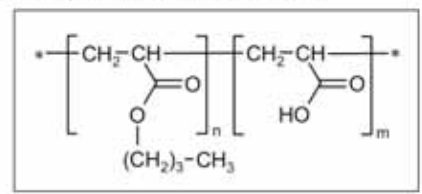

**Figure 1** : Chemical structure of the diblock copolymer PBA-b-PAA investigated in the present work. The abbreviation PBA stands for poly(n-butyl acrylate) and PAA for poly(acrylic acid).

The different homo- and co-polymers synthesized in that way were terminated by one xanthate group. It was checked that the remaining transfer agent at the chain terminus did not alter the micelle formation in water, nor the phase behavior in mixed solvents. After polymerization, the solvent was removed by evaporation or by freeze-drying. The polydispersity of blocks and homopolymers was estimated to be 1.5 by gel permeation chromatography (GPC) [23,26].

| polymer | $N_{BA}$ | $N_{AA}$ |
|---|---|---|
| h-PBA$_{1K}$ | 8 | - |
| h-PBA$_{8K}$ | 62 | - |
| h-PAA$_{8K}$ | - | 111 |
| PBA$_{3K}$-b-PAA$_{12K}$ | 23 | 167 |
| PBA$_{8K}$-b-PAA$_{8K}$ | 62 | 111 |

**Table I** : List of homopolymers and copolymers synthesized in this work. The molecular weights of the butyl acrylate and acrylic acid monomers were taken at 128.17 g mol$^{-1}$ and 72.06 g mol$^{-1}$, respectively.

In the present study, emphasis will be on the homopolymers h-PBA1K, h-PBA8K and h-PAA$_{8K}$ as well as on the diblocks PBA$_{3K}$-b-PAA$_{12K}$ and PBA$_{8K}$-b-PAA$_{8K}$ (Table I). These polymers were dissolved in water/THF mixtures for which the THF mass fraction $X = m_{THF}/(m_{THF} + m_{water})$ was varied from X = 1 (pure THF) to X = 0 (pure water). Typical polymer concentrations were in the range 0.1 – 1 wt. % for copolymers and in the range 1 – 25 wt. % for the homopolymers. The samples were kept at room temperature under gentle steering for three days before measurements.

At this point, it should be mentioned that the phase behaviors for $PBA_{7.5K}$-b-$PAA_{7.5K}$ and $PBA_{12K}$-b-$PAA_{3K}$ in aqueous and polar solvents were reported recently [23]. In this work, it was found that block copolymer vesicles could be produced and isolated using film rehydration techniques and specific solvent conditions. For $PBA_{7.5K}$-b-$PAA_{7.5K}$, the thickness of the aggregates could be continuously adjusted in the range 50 – 500 nm by addition of poly(n-butyl acrylate) homopolymers prior to the film formation and rehydration processes.

## *II.2 – Experimental Techniques*

### II.2.1 – Dynamic Light Scattering

Static and Dynamic Light scattering was performed on a Brookhaven spectrometer (BI-9000AT autocorrelator) for the measurements of the scattering intensity and of the normalized second order autocorrelation function. A Lexel continuous wave ionized Argon laser was operated at the wavelength $\lambda$ = 514.5 nm and at incident power comprised between 20 mW to 150 mW. The sample temperature was controlled using an Haake thermostatic bath system between 20° C to 65° C with an accuracy of ± 0.1° C. The normalized first-order autocorrelation function $g^{(1)}(t)$ was derived through the relationship [36,37] :

$$g^{(2)}(t) = 1 + \beta \left| g^{(1)}(t) \right|^2 \quad (1)$$

where $g^{(2)}(t)$ is the normalized second order autocorrelation function, $\beta$ is a parameter of the optical system (constant) and t is the delay time. $g^{(1)}(t)$ was analyzed afterwards in terms of a continuous sum of exponential decays :

$$g^{(1)}(t) = \int_0^\infty G(\Gamma) \exp(-\Gamma t) \, d\Gamma \quad (2)$$

where $G(\Gamma)$ is the distribution of decay rates $\Gamma$. For block copolymers in mixed solvents, it was found that Eq. 2 could be appropriately replaced by a discrete sum of single exponential functions, each of them being characterized by a decay rate $\Gamma_i$ [38]. For selected samples, the decay rates $\Gamma_i$ retrieved from the fitting procedure were plotted against the scattering wave-vector $q = \frac{4\pi n}{\lambda} \sin(\theta/2)$ in order to verify that the different relaxation modes were diffusive, *i.e.* $\Gamma_i = D_0^i q^2$. In the previous expressions, the index i stands for "mic" (micelles), "mso" (micron-sized objects) or "uni" (unimers), as explained later in the text. Here, n denotes the refractive index of the solution, $\theta$ the scattering angle (comprised between 30° and to 145°) and $D_0^i$ the diffusion coefficient of the ith mode. The hydrodynamic diameters were calculated according to the Stokes-Einstein relation, $D_H^i = k_B T / 3\pi \eta D_0^i$, where $k_B$ is the Boltzmann constant, T the temperature (T = 298 K) and $\eta$ the solvent viscosity. The viscosity of the water/THF mixtures at 25° C was extrapolated from the viscosities of the pure systems, $\eta_{water}$ = 0.89 mPa s and $\eta_{THF}$ = 0.46 mPa s. In these systems, dynamic light scattering was operated in the dilute regime in order to minimize the effects of interactions on the diffusion coefficients [39].

### II.2.2 – Small-Angle Neutron Scattering

Small-angle neutron scattering was performed in aqueous solutions in order to ascertain the microstructure of the PBA-b-PAA polymeric micelles. The experiments were made at the Laboratoire Léon Brillouin (LLB, Saclay, France) and at the Institute Laue-Langevin (ILL, Grenoble, France) and the results were found to be consistent with each others. Here, only data obtained on the D22 beam line at the ILL are shown. On D22, the scattering cross-sections $d\sigma/d\Omega(q,c)$ were collected at sample-to-detector distances 2 m and 14 m, with an incident wavelength of 12 Å and a wave-vector resolution $\Delta q/q$ of 8 %. This configuration allowed to cover a wave-vector range comprised between $1.7 \times 10^{-3}$ Å$^{-1}$ and 0.24 Å$^{-1}$. $PBA_{3K}$-b-$PAA_{12K}$ and $PBA_{8K}$-b-$PAA_{8K}$ solutions were studied as a function of the concentration at room temperature (T = 25° C) and neutral pH (pH 8). At this pH, approximately 75 % of the acrylic acid monomers were ionized, resulting in a condensation of the sodium counterions and in a strong stretching of the outer chains. The coherent scattering length densities of the butyl acrylate and sodium acrylate monomers were estimated respectively at $\rho_N(BA)$ = + 0.56×10$^{10}$ cm$^{-2}$ and $\rho_N(NaA)$ = + 4.4×10$^{10}$ cm$^{-2}$, where NaA stands for sodium acrylate [27,40]. From these two values, it appeared that with respect to D$_2$O ($\rho_N(D_2O)$ = + 6.4×10$^{10}$ cm$^{-2}$) the poly(n-butyl acrylate) has a much better contrast than poly(sodium acrylate). We thus anticipate that the cores of the micelles should contribute predominantly to the total neutron scattering cross-section. The spectra were treated according to the ILL and LLB standard procedures, yielding neutron scattering cross-sections expressed in cm$^{-1}$.

### II.2.3 – Optical Microscopy

Differential interference (DIC) and phase contrast microscopy was carried out to visualize micron-size objects using 40× and 100× magnifications. Images were monitored with a color CCD camera and stored by a computer. Fluorescent microscopy using fluorescent salts was also applied in order to assess the partitioning of the solvents induced by polymers in the different phases. To this aim, we have utilized Alexa Fluor® 488, which is a green fluorophore and Alexa

Fluor® 594 that emits in the red. Both are water soluble salts but they are insoluble in THF.

II.2.3 – Nuclear Magnetic Resonance

The instrument used is UNITY INOVA 400MHz with a 4 nuclei switchable probe. Samples were dissolved in deuterated DMSO for each $^1$H-NMR experiment. We varied on the concentration of the polymer to obtain a good integration. In addition, to insure accuracy of the water signals in such low concentrations of THF, many repetitions were made. We also subtracted water signals coming from DMSO alone. Moles ratios of solvent to water were calculated based from these NMR results.

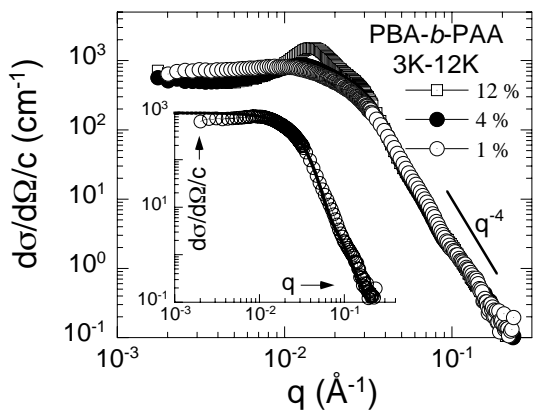

**Figure 2** : *Small-angle neutron scattering cross-sections obtained on PBA$_{3K}$-b-PAA$_{12K}$ aqueous solutions at concentrations 1, 4 and 12 wt. %. The intensity was divided by the concentration in order to stress the superposition of the data in the high q-range. Inset : neutron scattering spectrum at c = 1 wt. % representing the form factor of the micellar aggregates. The continuous line is the result of best fit calculations using a model of polydisperse spheres (median diameter D$_C$ = 11.6 nm, polydispersity s = 0.25).*

## IV – Results and Discussion

### *III.1 – Micelle Formation in Aqueous Solvent*

Figs. 2 and 3 display the scattering cross-sections divided by the concentration $d\sigma/d\Omega(q,c)/c$ obtained for PBA$_{3K}$-b-PAA$_{12K}$ and PBA$_{8K}$-b-PAA$_{8K}$ solutions at concentrations ranging from 0.4 to 20 wt. %. For both systems, the cross-sections exhibited a strong forward scattering as $q \rightarrow 0$ with a clear saturation plateau, whereas the high q-range was dominated by a decrease of the intensity in the form of a power law $d\sigma/d\Omega(q) \sim q^{-n}$, with n ~ 4 (indicated by a straight line in these figures). As the concentration was increased, the spectra were found to superimpose at large wave-vectors. At low q however, and for concentration larger than 10 wt. %, a structure peak showed up around $10^{-2}$ Å$^{-1}$, resulting in a reduction of the overall scattering in this region.

The scattering features of Figs. 2 and 3 are representative of dispersed systems characterized by length scales in the nanometer range. These scattering features have been observed repeatedly by small-angle scattering during the last decades on various polymer and colloid systems [37] and are interpreted as follows [24,39,41-47] : i) in the dilute regime of concentration, *i.e.* where inter-aggregate interactions are negligible, the scattering cross-section represents the form factor of the aggregates. ii) the existence of a $q^{-4}$-power law at large q defines the Porod regime and ascertains the presence of sharp interfaces between the elementary scatterers and the solvent; iii) the fact that the intensity divided by the concentration superimposes in the Porod regime whatever c is a strong indication that the size and microstructure of the aggregates remain unchanged for these dispersions.

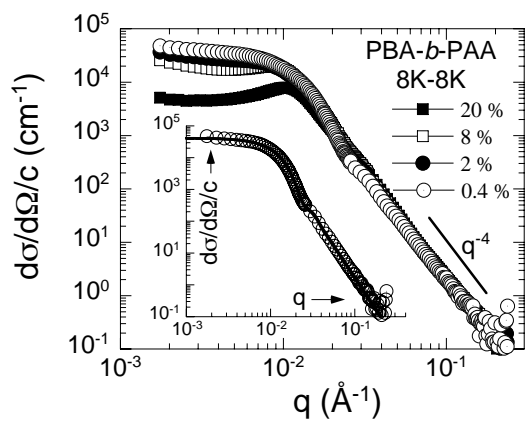

**Figure 3** : *Same as in Fig. 2, but for PBA$_{8K}$-b-PAA$_{8K}$ solutions. The continuous line in the inset was obtained using D$_C$ = 32.0 nm and s = 0.2.*

The form factors are emphasized in the insets of Figs. 2 and 3, together with fits based on a polydisperse sphere model (continuous lines). In the present case, the size distribution was assumed to obey a log-normal function distribution which characteristics are a median diameter D$_C$ and a polydispersity s. The polydispersity index s is defined as the ratio between the standard deviation and the average diameter. The formalism of the polydisperse sphere model can be found in numerous reports [37,40] and we refer to them for further details. As shown in the insets, the agreement between the data and the calculations is excellent for the two polymers. For PBA$_{3K}$-b-PAA$_{12K}$, one gets D$_C$ = 11.6 nm and s = 0.25, whereas for PBA$_{8K}$-b-PAA$_{8K}$ D$_C$ = 32.0 nm and s = 0.20. Considering that the scattering contrasts of the PBA and PAA blocks with respect to D$_2$O are very different, ($\Delta\rho_N$(BA) = 5.84×10$^{10}$ cm$^{-2}$ and $\rho_N$(NaA) = + 2.0×10$^{10}$ cm$^{-2}$), and that the overall intensity is proportional to $\Delta\rho_N^2$, it can be assumed here that the scattering arises essentially from the poly(n-butyl acrylate) chains, *i.e.* from the cores of the micelles. The size distributions obtained by fitting represent then the distributions of the core sizes of the micelles. This assumption is further supported by i) the values of the radius of gyration as estimated from the low-q region ($R_G$ = 6.8 nm and 16.3 nm) and ii) from the values of the hydrodynamic diameters obtained from dynamic light scattering measurements. For $PBA_{3K}$-b-$PAA_{12K}$ and $PBA_{8K}$-b-$PAA_{8K}$, $D_H$ is of the order of 100 nm, indicating moreover a significant stretching of the acrylic acid blocks (Table II) [48,49]. The average number of diblocks per micelle $N_{Agg}$ can be inferred from the core size distributions, through the relationship :

$$N_{Agg} = \frac{\pi \overline{D^3}}{6 N_{BA} v_0} \qquad (3)$$

where $\pi \overline{D^3}/6$ is the average volume of the core, $N_{BA}$ the number of repeating units and $v_0$ the molecular volume of a monomer. Using $v_0$ = 268 Å$^3$, we have found $N_{Agg}$ = 200 for $PBA_{3K}$-b-$PAA_{12K}$ and $N_{Agg}$ = 1400 for $PBA_{8K}$-b-$PAA_{8K}$. These values, as well as those of the core diameter and corona thickness are in good agreement with data reported recently on similar PBA-b-PAA systems (Table II) [24,26,27].

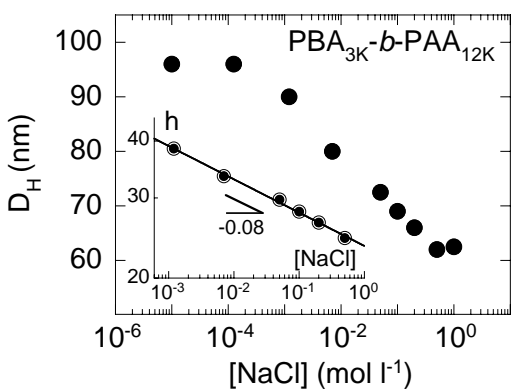

***Figure 4*** *: Hydrodynamic diameter measured as a function of the sodium chloride molar concentration for $PBA_{3K}$-b-$PAA_{12K}$ block copolymers in water (pH 8). Inset : scaling of the shell thickness with the salt concentration (h ~ [NaCl]$^{-0.08}$).*

Fig. 4 shows the evolution of the hydrodynamic diameters for $PBA_{3K}$-b-$PAA_{12K}$ as a function of added NaCl in water. With increasing [NaCl] concentration, $D_H$ first remained constant at a value of the order of 100 nm, and above [NaCl] = 10$^{-4}$ – 10$^{-3}$ M, it exhibited a cross-over toward a regime where the hydrodynamic sizes of the micelles shrunk gradually. This behavior is again typical for core-shell colloids where the corona is made of charged polymers. Theoretically, it has been described in terms of the contraction of the spherical brush that results from the screening of the electrostatic interactions. Based on a blob model developed for polyelectrolytes, Hariharan et al. have proposed that the brush thickness $h$ = ($D_H$ – $D_C$)/2 scales with the salt concentration as $h$ ~ [NaCl]$^{-m}$, where m is an exponent that depends on the curvature of the tethered surface. m was predicted to be 1/3 for zero curvature substrate and m = 1/10 for highly curved surfaces. The straight line in the inset of Fig. 4 was obtained assuming such a power law dependence with m = 0.08. The exponent is close to that observed for poly(styrene sulfonate) [50-52] and for poly(acrylic acid) [8] brushes. The data in Fig. 4 finally confirm that without added salt the polymers in the corona are strongly stretched since $h$ is found of the order of the contour length of the poly(acrylic acid) block. In conclusion of this section, we have found that in water, PBA-b-PAA block copolymers self-assemble into core-shell micelles and that the microstrucutre evidenced by light and neutron scattering is in good agreement with that reported in the literature on the same systems.

| polymer | DLS | SANS | | | |
|---|---|---|---|---|---|
| | $D_H$ (nm) | $R_G$ (nm) | $D_C$ (nm) | s | $N_{Agg}$ |
| $PBA_{3K}$-b-$PAA_{12K}$ | 96 | 6.8 | 11.6 | 0.25 | 200 |
| $PBA_{8K}$-b-$PAA_{8K}$ | 88 | 16.3 | 32.0 | 0.20 | 1400 |

**Table II** : List of parameters derived from dynamic light scattering (DLS) and small-angle neutron scattering (SANS) experiments performed PBA-b-PAA aqueous solutions ($D_H$ : hydrodynamic diameter; $R_G$ : radius of gyration; $D_C$ : median diameter; s : polydispersity; $N_{Agg}$ : aggregation number). Experiments were conducted at room temperature and neutral pH.

### III.2 – h-PAA and h-PBA homopolymers phase behavior in mixed solvents

In order to interpret the behavior of PBA-b-PAA in mixed solvents, the study of the homopolymers, h-PBA and h-PAA, were first carried out. It is known that water and THF are both good solvents for the acrylic acid monomers [21]. This property was verified on a 8000 g mol$^{-1}$ h-PAA that was synthesized for the present work. Similarly, this homopolymer was found to dissolve readily in mixed water/THF solvent whatever the mixing ratio X.

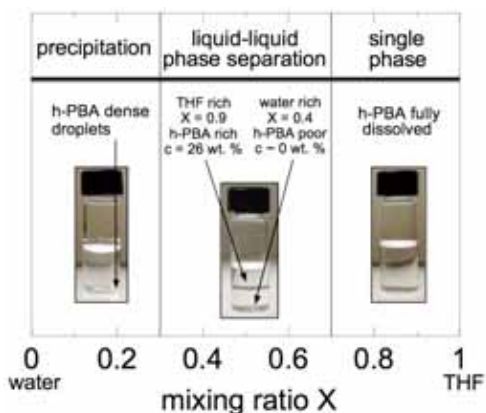

***Figure 5*** *: Schematic phase diagram for poly(n-butyl acrylate) (h-PBA) in mixed THF/water solvents. The mixing ratio X is the THF mass fraction in the solvent. At*

*intermediate X (0.3 ⩽ X ⩽ 0.7), a liquid-liquid phase separation is observed.*

Poly(n-butyl acrylate) homopolymer presented a different phase behavior. At c(h-PBA) = 5 wt. % and high THF mass fractions *i.e.* for X > 0.7, we have observed that 1000 g mol$^{-1}$ and 8000 g mol$^{-1}$ polymers could be dissolved easily, yielding transparent solutions. On the other side of the mixing diagram, when the water fraction becomes the most important (X < 0.3), h-PBA precipitated at the bottom of the container under the form of a melt. At intermediate mixing ratios, *i.e.* for 0.3 ⩽ X ⩽ 0.7, the system exhibited a liquid-liquid phase separation. Fig. 5 illustrates the phase behavior of h-PBA in mixed water/THF solvents for X comprised between X = 0 and X = 1. This transition is further illustrated in Fig. 6 where the proportions of the upper phase are displayed as a function of the initial polymer concentration. All samples were prepared at X = 0.7. For both molecular weights (M$_W$ = 1000 g mol$^{-1}$ and 8000 g mol$^{-1}$), the proportions were of the same order and decreased linearly with concentration. This result indicates that for low molecular weight chains, the degree of polymerization of the polymer does not influence the liquid-liquid phase separation.

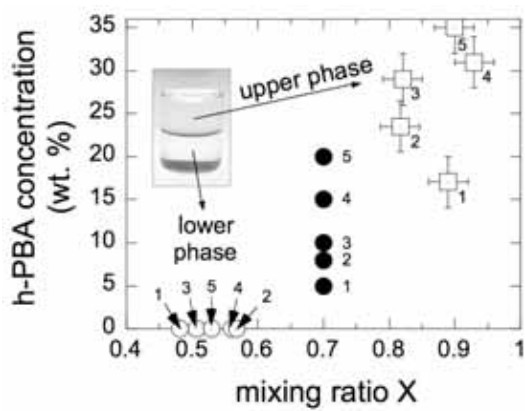

***Figure 7*** : *Liquid-liquid phase separation of h-PBA in mixed THF/water solvents : h-PBA1K concentrations versus mixing ratio for the upper and lower phases investigated separately using 1H-NMR. The initial samples labeled 1 to 5 were prepared at the initial concentration c(h-PBA) = 5, 8, 10, 15, 20 wt. % respectively and mixing ratio X = 0.7. The data show that h-PBA chains in mixed water/THF induce a partition of the solvent.*

Analysis of the different phases were subsequently performed using 1H-NMR for samples prepared at X = 0.7 and concentrations comprised between c(h-PBA) = 2 wt. % and 25 wt. % (h-PBA, MW = 1000 g mol$^{-1}$). By integration of the water THF and peaks from the NMR spectra, we have estimated the mixing ratio of each phase, noted X$_{up}$ and X$_{low}$ in the following. Similarly, the polymer content was obtained from the integration of the butyl acrylate hydrogen signal, yielding the quantities c$_{up}$(h-PBA) and c$_{low}$(h-PBA) for the upper and lower phases. Fig.7 shows the c(h-PBA) versus X phase diagram for the different upper and lower phases collected (initial samples are labeled 1 to 5 in the figure, corresponding to c(h-PBA) = 5, 8, 10, 15, 20 wt. % respectively). The upper solutions were found to be rich in THF, with an internal mixing ratio around X$_{up}$ = 0.9 ± 0.03 and a h-PBA concentration of the order of c$_{up}$(h-PBA) = 25 wt. %. Note the slight increase of the polymer concentration as one passes from sample 1 to sample 5, and also the consistency of this increase with respect to the data displayed of Fig. 5. The solutions in the lower phase were found to have an internal mixing X$_{low}$ = 0.53 ± 0.04 and a polymer content below the NMR-detection limit, *i.e.* c$_{low}$(h-PBA) = 0 ± 2 wt. %. 1H-NMR performed on the 8000 g mol$^{-1}$ h-PBA have revealed similar results. The main findings of this study is thus that hydrophobic poly(n-butyl acrylate) chains in mixed water/THF induce a liquid-liquid phase separation associated with a partition of the solvent. The partition yields a THF-rich phase (X = 0.90) that contains most of the polymers and a phase at intermediate mixing ratio (X ~ 0.5) devoid of polymers. In the next section, we investigate the role of the adjunct of a hydrophilic poly(acrylic acid) block on the h-PBA chain.

### III.3 – PBA-b-PAA block copolymers phase behavior in mixed solvents

The phase behavior of PBA-b-PAA block copolymers in mixed solvents has been studied using the same protocols as those applied for the homopolymers. The two copolymers PBA$_{3K}$-b-PAA$_{12K}$ and PBA$_{8K}$-b-PAA$_{8K}$ were investigated. The main result of the study was that with copolymers, the solutions remained monophasic whatever X comprised between 0 and 1. The liquid-liquid phase separation seen with h-PBA and illustrated in Figs. 6 and 7 was not observed with none of the two diblocks. In the following, we describe dynamic light scattering data obtained on the PBA$_{3K}$-b-PAA$_{12K}$ system in the dilute regime. Fig. 8 shows the normalized first-order autocorrelation function g(1)(t/η) obtained at c(PBA$_{3K}$-b-PAA$_{12K}$) = 0.6 wt. % and X = 0.3, 0.6, 0.7 and 0.8. In order to take into account the effects of mixing, the delay time in abscissa has been normalized by the water/THF viscosity.

At high THF contents (X > 0.75) the autocorrelation functions were characterized by a single exponential decay related to a unique class of objects of diameter $D_H^{uni}$ = 5 ± 1 nm (average taken on the different X-values in this range). The D$_H$'s were calculated from the Stokes-Einstein relationship and plotted in Fig. 8 as a function of X. In this region, we recall that the two blocks taken separately are soluble. It can thus be assumed that in this range, light scattering arises from unassociated PBA$_{3K}$-b-PAA$_{12K}$ unimers.

With addition of water (0.4 ≤ X ≤ 0.75), the solutions were found to undergo a transition from transparent to bluish or slightly scattering, corresponding to a steep increase of the scattered light [6,30,33]. Autocorrelation functions $g_{(1)}(t)$ revealed two or three distinctive relaxation modes, each being separated from the others by one decade in decay time or so. As already mentioned in the experimental section, $g^{(1)}(t)$ was interpreted in terms of a sum of single exponential functions, with decay rate noted $\Gamma_i$. Details about the procedures and the reliability of the fitting are provided in the Supporting Informations Section. It was furthermore verified that the three modes were all diffusive by studying the wave-vector dependence of the decay rates. For scattering angles comprised between 30° and 140°, we found a $q^2$-variation for the $\Gamma_i$'s, allowing the determination of the diffusion coefficients $D_0^i = \Gamma_i(q)/q^2$ and hydrodynamic diameters $D_H^i$ for each mode. Observed at higher THF content, the fastest mode around $D_H^{uni}$ = 5 nm was found to persist in the range 0.5 ≤ X ≤ 0.75 (Fig. 9). By continuity with the regime at high THF mass fraction, it was then attributed to the unassociated unimers.

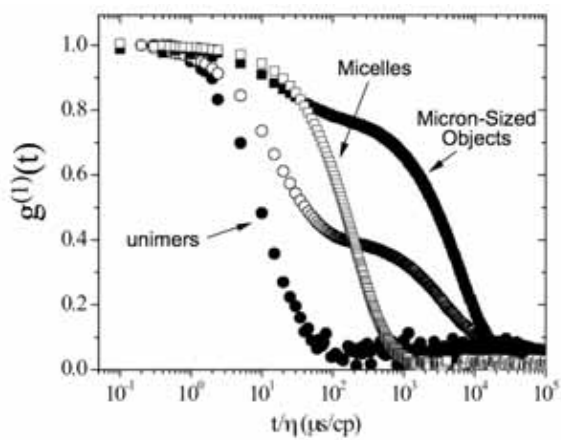

***Figure 8*** *: Normalized first-order autocorrelation function $g^{(1)}(t/\eta)$ obtained at for $PBA_{3K}$-b-$PAA_{12K}$ block copolymers at concentration c = 0.6 wt. % and mixing ratios X = 0.3 (empty squares), 0.6 (close squares), 0.7 (empty circles) and 0.8 (close circles). The delay time in abscissa has been normalized by the water/THF viscosity in order to compare relaxation modes independently of the solvent viscosity. The contribution of the scattered light arising from unimers, micron-sized objects and micelles are indicated.*

The slowest relaxation mode in Fig. 8 was found to be associated with large scattering objects of diameters 0.5 – 2 µm. Observations by optical microscopy have confirmed the presence of micron-sized objects of spherical symmetry with sizes comparable to those seen by light scattering. In the following, these large structures will be referred to as micron-sized objects or "mso". With X decreasing down to 0.4, $D_H^{mso}$ was found to decrease from ~ 1000 nm to around 400 nm at the boundary limit. Simultaneously, an additional third mode developed at intermediate decay times, associated with sizes around 100 nm. At X = 0.5, the three modes described previously could be observed on the same spectrum. Their 's are represented in Fig. 9.

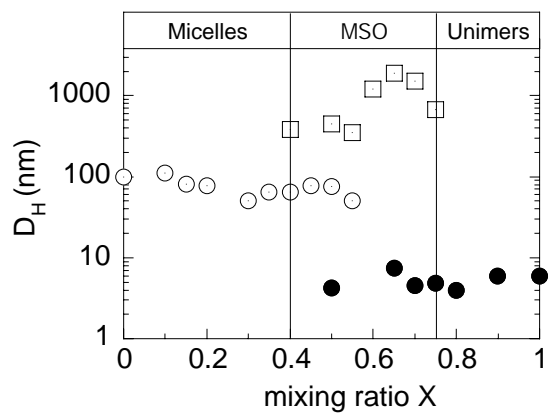

***Figure 9*** *: Evolution of the hydrodynamic diameter as a function of the mixing ratio X between water and THF. The $PBA_{3K}$-b-$PAA_{12K}$ concentration is 0.6 wt. % for all samples. The phase diagram exhibits three distinctive $D_H$-ranges, noted "Micelles", "Micron-Sized Objects" and "Unimers".*

For mixing ratio X < 0.4, the autocorrelation functions were characteristized by a unique relaxation mode, with sizes $D_H^{mic}$ = 60 – 100 nm. By continuity with the case of water, we conclude that in this interval, the mode seen by light scattering originates from micellar aggregates similar to those described in Section III.1.

In conclusion, we have found that with block copolymers, mixed solutions did not exhibit a phase separation. In addition to the expected unimers and micellar phases, a dispersion of micron-sized spherical objects could be evidenced. It is important to emphasize here that this phase showed up in the same interval where h-PBA exhibited a liquid-liquid phase separation.

### III.4 – Nature of the Micron-Sized Objects

Among the different investigations of block copolymers in mixed solvents, Eisenberg and coworkers have reported the existence of thermodynamically stable vesicles [9,10,30] when polystyrene-b-poly(acrylic acid) block copolymers were dissolved in solutions of dioxane/THF/water or DMF/THF/water. The mechanism for thermodynamic stabilization of vesicles was described as a preferential segregation of the copolymers within the membrane, with short hydrophilic blocks migrating to the inner layer of the vesicle membrane and diblocks with long PAA chains to the outer [10]. Since the sizes of the equilibrium vesicles found with PS-b-PAA matched approximately those of the large objects identified in Fig. 8, the question about their structure and composition was addressed. Are the micron-sized objects found with PBA-b-PAA vesicles similar to the ones reported in the literature [23,30,33], or are they the result of a partitioning of the solvent mixture, as the experiments on h-PBA could suggest it ? In the first assumption, a vesicle would enclose an internal phase identical to the outer one, the separation between the two being a single or a stack of copolymer bilayers. In the second assumption, these micron-sized objects would be droplets of one phase dispersed in a medium of another phase.

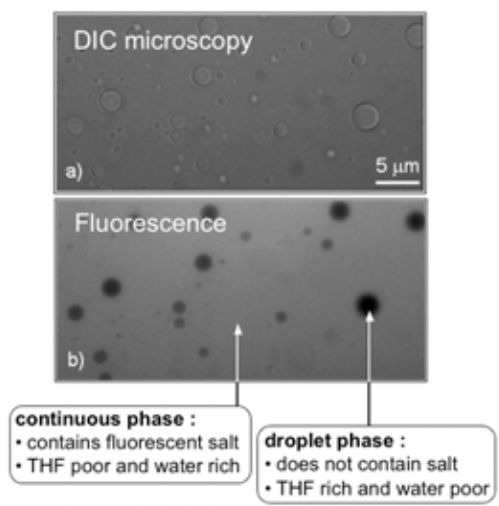

Figure 10 : Droplet phase of $PBA_{3K}$-b-$PAA_{12K}$ dissolved in THF/water solvent at c = 1 wt. % and X = 0.6. The mixed solvent contains $10^{-5}$ mole $L^{-1}$ of a fluorescent salt (Alexa Fluor 594) soluble in water and insoluble in THF. a) Upper panel : image of 20×40 $\mu m^2$ field observed by differential interference contrast microscopy (DIC). b) lower panel : image of the same field observed by fluorescent microscopy. The 1 µm-droplets appeared to have a different fluorescence contrast with respect to the background. The fluorescent continuum represents the water-rich region and the non-fluorescent droplets (dark) are THF-rich.

We have used fluorescent microscopy in order to elucidate the nature of the dispersion of droplets found with PBA-b-PAA at intermediate water/THF mixing ratio. High molecular weight ionic species are known to be soluble in water and insoluble in THF. This property makes high molecular weight salts interesting to differentiate vesicles from THF rich droplets in water. Monovalent fluorescent salt (Alexa Fluor 594, excitation maximum at 580 nm and emission maximum at 610 nm) was dissolved in water at a concentration of $10^{-5}$ molar. THF was then added in order to achieve the final mixing ratio of X = 0.6. The solution appeared transparent and the fluorescence was found to be uniformly distributed on a 100 nm length scale, indicating a good mixing of the two fluids. When $PBA_{3K}$-b-$PAA_{12K}$ was added to the previous X = 0.6 solvent at c($PBA_{3K}$-b-$PAA_{12K}$) = 1 wt. %, a dispersion analogous to the one reported in the previous section formed spontaneously. It was confirmed by dynamic light scattering that the three relaxation modes corresponding to unimers, micron-sized objects and micelles were present.

Fig. 10 displays two images of the same field, obtained by differential interference contrast microscopy (DIC, upper panel, Fig. 10a) and by fluorescent microscopy (lower panel, Fig. 10b). The field covered on the viewgraph is approximately 20×40 $\mu m^2$ and was obtained with a 40×magnification. As a result, the droplets observed in Fig. 10a were characterized by a median diameter of $D_H^{mso} \sim 1$ µm. More importantly they appeared to have a different fluorescence contrast with respect to the background continuum (Fig. 10b). The inside of the droplets (in dark) did not fluoresce but the continuum did. As the inner and the outer phases of the droplets appeared different, the vesicular nature of the dispersion can be ruled out. Moreover, because of the specific solubility properties of the Alexa Fluor 594 salt, Fig. 10 suggests that fluorescent continuum represent water regions rich, and the non-fluorescent droplets are THF rich.

The picture that can be derived from the previous observations is consistent with the h-PBA homopolymer phase behavior in mixed water/THF. We have found that the copolymers have induced a microphase separation. Thanks to the data collected on h-PBA (Fig. 6 and 7), we anticipate that the droplets should be concentrated in polymers, in a solvent dominated by THF. The main difference between the homo- and block copolymers is that the h-PBA driven phase separation is arrested at a micron scale by the poly(acrylic acid) blocks. In this particular case, it can be assumed that at the interface between the droplets and the majority phase, the PAA blocks are threading towards the continuum phase and stabilize the droplets. Note finally that at X ~ 0.7, the dynamic light scattering experiments are probing both unimers and droplets (at this stage polymeric micelles are not fully formed), but here the unimers are inside the droplets. In terms of translational diffusion time scale, both phenomena are decoupled and can be observed by this technique.

The stabilization of the droplet dispersion was found to be strongly dependent on the molecular weight, as well as on the PBA/PAA relative mass ratio. The phase behavior of the symmetrical diblock $PBA_{8K}$-b-$PAA_{8K}$ in mixed solvents was obtained and it yielded globally the same results as for the $PBA_{3K}$-b-$PAA_{12K}$ : at intermediate X (0.4 < X < 0.8), larger droplets in coexistence with unimers or micelles were formed. With this polymer, however, the drop sizes were larger than those obtained with PBA$_{3K}$-b-PAA$_{12K}$ ( ~ 10 μm). As the result of their coalescence, creaming and eventually a macroscopic phase separation occurred over time. These findings suggest again that the molecular weight of the PAA block plays a role in the determination of the average droplet curvature, and in fine in the stabilization of the dispersion.

### III.5 – Kinetically frozen versus equilibrium states : effect of dilution and temperature

Block copolymers in aqueous solvents are usually assumed to have a low or very low critical micellar concentration [1,26,53]. We have used static light scattering in order to determine the cmc of the PBA$_{3K}$-b-PAA$_{12K}$ and PBA$_{8K}$-b-PAA$_{8K}$ systems in water (X = 0). With decreasing concentration, the scattering intensity expressed in terms of the Rayleigh ratios exhibited a linear dependence down to $10^{-4}$ wt. % [54]. Below this value, the scattering intensity became of the order of the error bars of the measurements. This linear dependence is in agreement with the dilution law of colloidal systems, which indicates that the scattering constituents remain unaltered by the dilution process. Moreover, in the same concentration range, dynamic light scattering performed on PBA$_{3K}$-b-PAA$_{12K}$ revealed the persistence of a single relaxation mode with hydrodynamic diameter $D_H^{mic}$ = 100 nm. X-ray scattering experiments performed as a function of the temperature (data not shown) have demonstrated that the micelles retained their microstructure between T = 20° C and 60° C [53]. All these findings suggest that the cmc for PBA-b-PAA remains below $10^{-4}$ wt. % [26], and that although the glass transition temperature for PBA is below room temperature ($T_G$ = - 52° C), the structure of the polymeric micelles in water can be considered as kinetically frozen. These results are in contradiction with recent experiments by Gaillard and coworkers [22] who have evaluated cmc's for PBA-b-PAA with total molecular weights comprised between 2000 and 6000 g/mol and different mass ratios in the range cmc(PBA-b-PAA) = $10^{-1}$ – $10^{-3}$ wt. %. In particular, for a system with a 3000 g/mol PBA block, that is with an hydrophobic block comparable to that of the PBA$_{3K}$-b-PAA$_{12K}$ studied here, these authors found a cmc of 0.2 wt. %. We have no explanation for the discrepancies between the two surveys.

In this context, it was then important to assess if the droplet dispersions found in mixed solvents at intermediate X was at thermodynamic equilibrium. In the experiment of Fig. 11, the THF/water ratio was fixed at X = 0.6, that is in the region of the phase diagram where unimers and droplets were both apparent in light scattering. Temperature loops consisting in a slow heating and a subsequent cooling of the system between 25 ºC and 65 ºC were carried out. The autocorrelation functions of the scattered intensity $g^{(1)}(t)$ obtained in the 90°-configuration are represented in Fig. 11 for T = 25, 45, 55 and 65° C.

As the temperature was increased, the relaxation mode associated with the diffusion of droplets vanished gradually. At the highest temperature (T = 65 ºC), it even disappeared completely. Meanwhile, the sample underwent a transition from turbid to transparent. When the temperature was reduced back to 25° C, the slow mode showed up again, with the same decay rate. From these data, we conclude that during the temperature loops, the unimer and droplet contributions coexist and the amplitudes related to the two relaxation modes vary reversibly. We have observed a similar behavior when the 1000 g mol$^{-1}$ h-PBA was dissolved at the same water/THF ratio and submitted to similar temperature ramps. At high temperatures, a single transparent phase was observed whereas at room temperature the sample was characterized by two coexisting phases. Moreover, when the temperature was increased, the volume of the upper phase diminished gradually and even disappeared at high temperature. Here again, the phase behavior of the block copolymers in mixed solvents mimics that of h-PBA in a very accurate way. These findings are a strong indication that the droplet phase for PBA-b-PAA in mixed water/THF solvent is at thermodynamic equilibrium, in contrast to the kinetically frozen micelles found in the water rich part of the phase diagram.

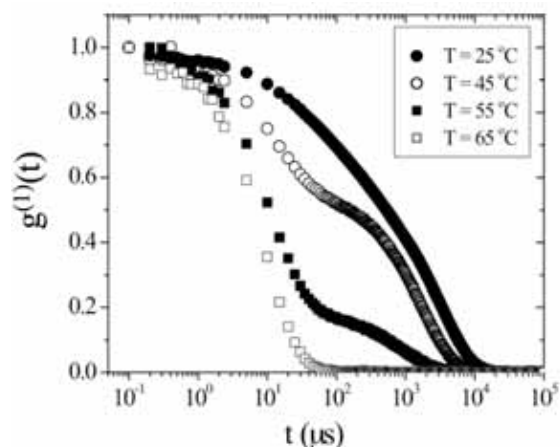

*Figure 11* : Normalized first-order autocorrelation function $g^{(1)}(t)$ obtained for PBA$_{3K}$-b-PAA$_{12K}$ block copolymer at concentration c = 1 wt. % and mixing ratios X = 0.6 at different temperatures between 25° C and 65° C. As the temperature increases, the relaxation mode associated to the droplets vanished gradually. The phenomenon is reversible.

## IV – Conclusion

The study of the phase behavior of PBA-b-PAA amphiphile block copolymers in mixed water/THF

solvents have revealed interesting new features. Starting from dispersed unimers in THF which is a good solvent for both poly(acrylic acid) and poly(butylacrylate), the progressive addition of water was found to induce a turbidity transition at a water mass ratio of 25 % (corresponding to X = 0.75 in Fig. 9). In contrast to the morphological transitions found in similar systems [30,33], for PBA-b-PAA we have evidenced a microphase separation. Based on fluorescent microscopy, the micro-separated state was described unambiguously as a dispersion of polymer and THF rich droplets in the 1 – 10 μm size range (Fig. 12). This state persisted down to a mixing ratio X = 0.4. In order to prove that the ternary system PBA-b-PAA/water/THF exhibited such a particular behavior, we have investigated h-PBA homopolymers with molecular weight 1000 and 8000 g mol$^{-1}$ under the same conditions. In the range where the microphase separation occurred, we have observed that h-PBA/water/THF exhibited a macroscopic liquid-liquid phase separation. NMR analysis of the different liquid phases have disclosed that one phase was rich in THF and contained most of the polymers, whereas the other was an equal mixture of the two solvents (X ~ 0.5) and scarce in polymers. From this comparison, it was concluded that for the diblocks, the phase separation and solvent partitioning induced by h-PBA was arrested at a micron scale thanks to the poly(acrylic acid) blocks that stabilized the droplets.

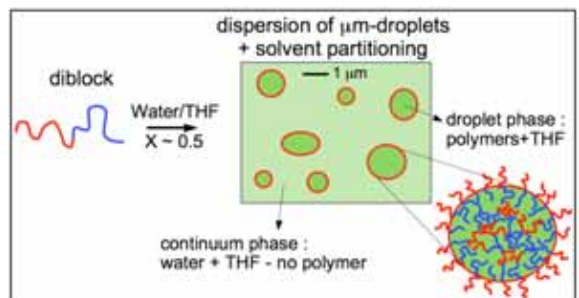

***Figure 12*** *: Schematic representation of the droplets phase obtained with PBA-b-PAA block copolymers in mixed water/THF solvents.*

It has been reported recently that although water and THF are two fully miscible fluids, the possibility of complexation of the cycloether molecules with water remains. Using dynamic light scattering, Yang et al. have shown that this complexation resulted in the formation of water-tetrahydrofuran clusters with typical sizes 0.3 – 1 nm, as well as large and unidentified domains in the range 200 – 600 nm [29]. In order to verify that the behavior evidenced in the present survey could not be attributed nor linked to these singular properties, PBA-b-PAA block copolymers were also studied in water/ethanol solvents for comparison. In water/ethanol, we confirmed the existence of a polymer induced micro-separated phase, again characterized by micrometer-sized droplets and a solvent partitioning. The data on the water/ethanol mixed systems will be published in a forthcoming publication. These results finally suggest that the phenomenon that consists in enclosing an organic solvent into large droplets by addition of copolymers (sponge effect) could be useful for applications, as e.g. in the solvent extraction industry.

**Acknowledgements** : The authors thank Rhodia for technical and financial support. O. Théodoly and D. Bendejacq are acknowledged for their comments on the manuscript.